# The Exotic Quasisolidity and Supersolidity of Water

Chang Q Sun[1*]

**Abstract**

Water is ubiquitously important but least known. This perspective features the latest finding of two exotic forms of water called quasisolid and supersolid phases due to the cooperativity and disparity of the O:H-O bond in its segmental length, energy, and specific heat when subjected to thermal, electric, and undercoordination perturbation. The quasisolid (QS) phase only appears at temperatures between melting and ice nucleation, regardless of sample size, and this phase has a negative thermal expansivity (NTE). However, polarization or molecular undercoordination creates the supersolid phase in skins of water and ice, droplets, ionic hydration cells, and volumetric water with a current flow or under a $10^6$ eV/cm electric bias. This phase is characterized by ~10% H-O bond contraction and vibrating at ~3450 cm$^{-1}$. The supersolid has unique properties such as high elasticity, mechanical strength, optical reflectivity, structure order, thermal stability, diffusivity, catalytic activity, and chemical reactivity. Liquid-QS-Ice transition with QS phase energy absorption occurs for water due to H-O bond contraction during thermal decay but not for the supersolid phase of saturated NaCl solution. The NTE of the QS fosters ice buoyancy and the supersolidity endows ice slipperiness, supercooling, superheating, premelting, low-temperature fiber elasticity, water bridge, and warm water fast cooling, as well as on water chemistry.

[1] RISE and School of Materials Science & Engineering, Dongguan University of Technology, Dongguan 523808, China

**TOC entry**

**The θ(t) decay lines revealed Liquid-QS-Ice transition and the QS phase energy absorption for water but not for the saturated NaCl supersolid solution.**

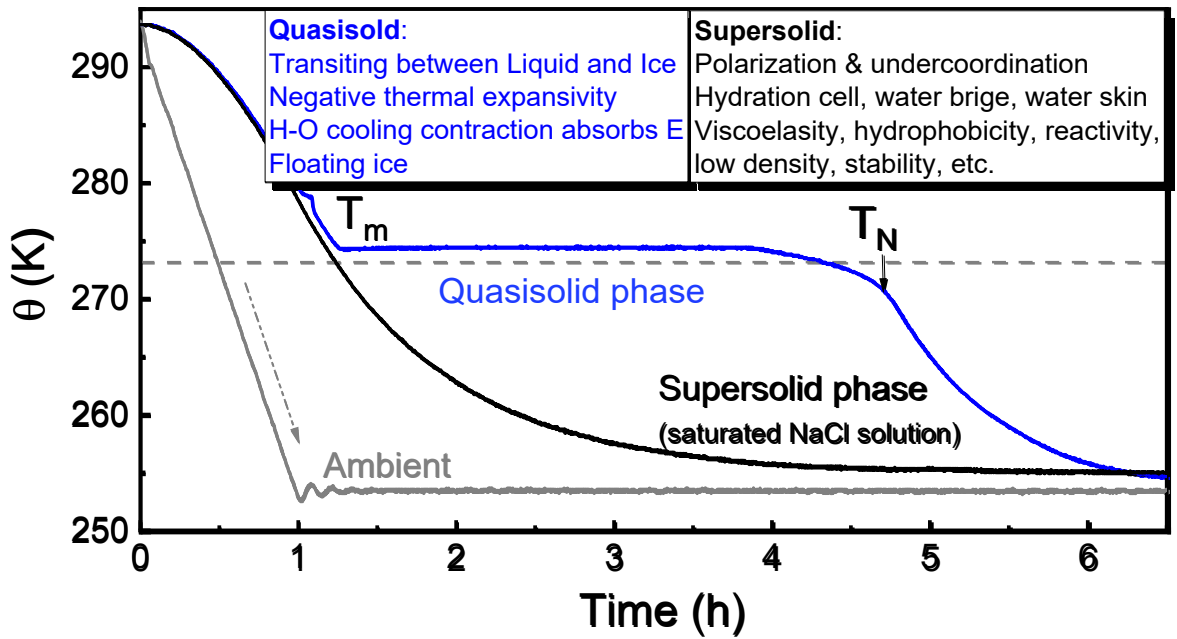

key learning points

✧ O:H-O bond specific-heat disparity entitles the quasisolity of negative thermal expansivity.

✧ Bound by $T_m$ and $T_N$, the quasisolid phase absorbs energy through H-O cooling contraction.

✧ Electrification or undercoordination creates supersolid shifting its 3200 cm$^{-1}$ to 3450 cm$^{-1}$.

✧ The H-O contracts ~10% in the supersolid skin of water ice, hydration cell, and water bridge.

✧ The viscoelastic, hydrophobic supersolid is less dense, more thermally diffusive, and stabiler.

About the principal author:

CQ, received a PhD from Western Australia in Physics in 1992, and has solo-authored four monographs published by Springer-Nature and their Chinese versions during 2014-2023, principally authored 35 perspectives/Reviews published in Chem Rev (2), Phys Rep. Prog Mater Sci (2) Surf Sci Rep, Coord Chem Rev (2), etc., and 500 peer-reviewed journal articles, with 20k citation hits, CQ was ranked top 184[th] among 246,856 Physicists in 2019 and remained within 1/1000 onward by academic impact.

# The Exotic Quasisolidity and Supersolidity of Water    1

## 1. Introduction

Water is most abundant but least understood. It is important to life on Earth but performs anomalously disobeying basic laws in physics. Water demonstrates extraordinary adaptivity, cooperativity, recoverability, reactivity, and sensitivity when subjected to a perturbation such as pressure, temperature, electrification, and molecular undercoordination.

Water shows at least 17 phases when subjected to changes in pressure and temperature. Under ambient pressure, water undergoes the XI-I-Liquid-Vapor phase transition and density oscillation. The phase boundaries also vary with the load of a perturbation. Compression lowers the melting point $T_m$ and raises the temperature for ice nucleation $T_N$, according to the phase diagram. The $T_m(P)$ drops by -22 °C under 210 MPa and rises by +6.5 °C under -95 MPa tension [1, 2]. In the liquid and ice-I, water follows the regular law of thermal expansion while at Liquid-Ice traction shows negative thermal expansivity [3-5].

Electrification or molecular undercoordination not only changes the $T_m$, $T_N$, and the evaporation point $T_V$ contrastingly to compression but also makes the anomalous water even more fascinating. For example, water skin is naturally the toughest, and ice is very slippery [6-10]. The surface stress of water reaches 72.75 mJ/m$^2$ at 293 K, which is higher than 26.6 mJ/m$^2$ of the CCl$_4$ solution. When a water droplet falls on water, it bounces around or rests on the surface before eventually disappearing, demonstrating the elasticity, hydrophobicity, and repulsivity of the contacting interfaces. Numerical duplication of the thermal effect on the surface stress resulted in the 182 K Debye temperature and the 0.09 eV cohesive energy for the O:H of the tetrahedrally coordinated $H_2O$ molecule at the surface [11].

As a specific shape of droplets, crystal ice fibers of sub-μm diameter are highly elastic and flexible at low temperatures [12, 13]. A 4.4 μm fiber has a bending elasticity of 10.9 % at 120 K in a high vacuum. Comparatively, bulk ice has the highest elasticity of 0.3% and is rigid and fragile. The ultralow pressure in high vacuum and tensile stress under e-beam radiation also lowers the $T_N$ of the fiber. The extraordinary elasticity demonstrates the coupling effect of multiple fields on the mechanical and thermal properties of the fiber. The fiber has a considerable volume of supersolid skin wrapping the fiber. High vacuum conditions, e-beam radiation, mechanical tension, and molecular undercoordination enhance one another to relax and polarize the HB, making the fiber more elastic and flexible.

Furthermore, water droplets confined within a hydrophobic pore or thin water films deposited on dielectric surfaces behave more like ice than liquid water at room temperature [14-20]. A water droplet can travel through a microchannel much faster than classical fluid theory would predict [21], while an air gap stays between the skin of the droplet and the wall. The hydrophobic contact creates a 3.8 Å thick skin of 0.71 g/cm³ density, in addition to a 0.6 nm air gap between the $SiO_2$ wall and the fluid [22, 23].

These phenomena are known as superfluidity, superlubricity, superhydrophobicity, and supersolidity (4S)[11], which share the common mechanism of being more elastic, repulsive, and frictionless while in relative motion. According to Wenzel-Cassie-Baxter law, a hydrophilic surface becomes even more hydrophilic and a hydrophobic one more hydrophobic when it is nanoscaled roughened [24]. The 4S phenomena of water droplets are crucial to many fields, such as transporting through hydrophobic channels for cell culturing and separating.

Droplets and bubbles of nanometer scales are more endurable [25-27], surviving days or even weeks compared to larger ones. The H-O bond of the skin and bulk contracts while the dangling one expands when heated. Skin H-O bonds of water and ice share the same of 3450 cm$^{-1}$ frequency, which are less sensitive to heating than the bulk one featured at 3200 cm$^{-1}$. [16, 28] Furthermore, the supersolid skin has a lower specific heat and high thermal diffusivity, which endows heat outward flow of the liquid in thermal transport of warmer water, called the Mpemba paradox [29].

Water droplets form ice at temperatures below the nucleation point $T_N$ and evaporate below the evaporation point $T_V$, which is named supercooling. Superheating happens below the melting temperature of $T_m$. The skin supersolidity endows the premelting of ice [30]. A droplet of 1.4 nm size becomes ice at 205 K [4], well below the bulk $T_N$ of 258 K [31]. Droplets of 3.4 and 4.4 nm across transit into ice at 220 and 242 K [32], respectively. In contrast, a 2.5 nm droplet melts at temperatures around 50 K above the $T_m$ of 273 K [33]. The monolayer water and the skin of bulk water remain as liquid till 325 K [34] and 310 K [35].

Using AFM, X-ray, and neutron reflectometry, James et al. [33] uncovered that a droplet grown ($\Delta m \geq 0$ mass increase) in a given humidity ambient on monolayer water film remains ice-like, hydrophobic till 323 K ($\Delta m = 0$). In contrast, evaporation occurs at 338 K ($\Delta m < 0$). The growing droplets exhibit superheating in melting $T_m = 273$ K and supercooling in evaporating, $T_V = 373$ K under ambient pressure. Similarly, a droplet having a higher surface curvature took 70 sec longer to freeze at 269 K [36].

Water can form a 15 nm-long bridge crossing over the edges of two containers under ~$10^6$ eV/cm bias. The bridge has a rubber-like elasticity of 16 MPa and remains stable before melting [37-39]. Ionic hydration by salt and acid solvation produces hydration cells of a few shells of molecules [40]. The hydration cells form sublattices to interlock regularly with the HB network of the solutions. The water bridge and hydration cells share the same H-O bond vibration frequencies around 3450 cm$^{-1}$.

Microdroplets smaller than 20 μm in diameter possess extraordinary chemical reactivity, which underscore their distinctive properties and potential for environmentally friendly production. The droplets could generate hydrogen peroxide ($H_2O_2$) without requiring catalysts or applying a voltage, and the rate of $H_2O_2$ production is proportional to the surface curvature [41, 42]. Accumulating evidence [43] solidifies the theoretical proposition of an ultrahigh electric field ($10^7$ V/cm) across the interface, sufficient to induce spontaneous redox reactions.

The air-water interfacial electric field could generate free electrons or hydroxyl radicals, transit decarboxylation of benzoic acid into phenol [44], methane oxidation to oxygenates [45], carbon dioxide and nitrogen to urea [46] and enhance nitrogen fixation by making ammonia from nitrogen[47]. The on-water catalysis is promising as it leads to substantial acceleration of reaction rate when insoluble reactants stirred in aqueous suspension [48, 49].

A coarse-grain model [50] suggests that the surface electric field reaches ~$1.6 \times 10^7$ V/cm. Density functional theory (DFT) computations [35] revealed that one surface oxygen ion has -0.048 more electrons than it used to be, providing an air-water interface electric field. It is common that hydrophobic materials possess surface-bound negative charge [51] because of polarization.

The remarkable gathering of surface charge enables electricity harvesting from water droplets [52-54]. A mere 100 μl water droplet impacting the surface from a 15 cm height generates over 140 V voltage, a 200 μA current, and a maximum power of 50 W m$^{-2}$, surpassing conventional droplet generator designs by several orders of magnitude. Introducing a superhydrophobic surface-based droplet electricity generator [55] can enhance the average electrical output and robustness of a device that works at low-frequency impinging droplets.

These properties become even more pronounced at the nanoscale, where the fraction of the undercoordinated molecules increases with the surface curvature. The intriguing phenomena of water droplets are associated with skin density loss. According to XRD measurements [56], the skin O—O separation is 5.9% longer than its bulk value of 2.8 (should be 2.7) Å or 15.6% density loss at 298 K in comparison to the 4.6% skin O—O length contraction of the liquid methanol.

Since 1859, when Faraday and Thomson [57, 58] first discovered that a liquid-like layer not only welds two blocks of ice but also makes ice slippery, known as ice regelation [31, 59], scientists have been working hard to understand the anomalies of water and ice. However, it is hard to attack the aforementioned mysteries from the perspective of the two-body hydrogen bond of H···O attraction dominance or molecular dynamics of the point-charged structural models such as the TIPnQ series of fixed charge, constant bond angle and bond length.

In order to cope with the discussed issues, we introduced the concepts of quasisolidity[60] and supersolidity[35]. We show that it is essential to define the segmental specific heat for the O:H-O bond thermodynamics[60] and to clarify electrification induced bond length and polarization[61]. We show that the combination of the hydrogen bond cooperativity and polarizability (HBCP) premise [62] with the bond order-length-strength correlation and nonbond electron polarization (BOLS-NEP)[63] notion, or the consequence of polarization, dictates the performance of undercoordinated water.

Inspired by the supersolidity of solid helium whose fragments experience elastic, repulsive, and frictionless contacting motion [64], the supersolid water refers to an exotic, gel-like phase formed when subjected to polarization by applying an electric field [61] or lowering the molecular number (CN) [62]. Charge injection by solvation [40], electric field induction [37-39], deposited on dielectric or charged surfaces [61, 65], entrapped in hydrophobic cavities, or by forming defects, skins, droplets, or bubbles.

Water CN changes with the surface curvature of an object in the $z_{cluster} < z_{droplet} < z_{plane} < z_{cavity} < z_{bulk} = 4$ order while molecules maintain their tetrahedral geometry regardless of the coordination environment. The extent of supersolidity increases with the surface curvature and the strength of the applied electric field. Tension, lower pressure, and electron beam radiation can enhance the supersolidity, while thermal loading destroys the supersolidity [62]. Supersolid water demonstrates fascinating properties, as addressed in **Table 1**.

**Table 1**. Exotic common features of supersolid water and characteristics of the skin supersolidity water droplets.[6, 9, 16, 24-28, 31, 48, 52, 53, 62, 66-85]

| Driving forces (Existing forms) | ✧ Electrification (Volumetric water bridge) <br> ✧ Ionic hydration (Salt and acid hydration cells) <br> ✧ Molecular under-CN (Skins of water ice and droplets) | |
|---|---|---|
| Principles | ✧ Multifield-driven HBCP for coupling O:H-O oscillator pair <br> ✧ BOLS-NEP for molecular undercoordination | |
| Key constituents | ✧ O—O repulsive coupling dictates HBCP under perturbation <br> ✧ Localized skin $H_2O$ dipole endows surface charge density <br> ✧ Soft and elastic O:H oscillator fosters elastic adaptivity | |
| | Properties | Facts |
| Structure order | 3/4 mass density <br> High structure order <br> Gel-like rheology (Viscoelastic) | 8% O—O expansion (10% H-O contraction & 20% O:H expansion) |
| Mechanical | Water skin toughness <br> Rubber-like 16 MPa elasticity <br> Elastic adaptivity and electric repulsivity (Hydrophobicity) | 250 cm$^{-1}$ H-O phonon blueshift from 3200 cm$^{-1}$; O:H frequency shifts from 200 cm$^{-1}$ down |
| Thermal | Lubricity (Ice slipperiness) <br> Supercooling at $T_V$ and $T_N$ <br> 30-60 K superheating at $T_m$ <br> High thermal diffusivity <br> High thermal stability <br> Low specific heat | HB cohesive energies shift from (0.2, 4.0) to (0.1, 5.0) eV <br><br> -0.048 e/molecule surface charge accumulation <br><br> Polarization hinders electron and phonon dynamics; Skin shell reflection fosters standing waves <br><br> O1s level entrapment from 536 to 538 eV; Nonbonding electron-bound energy reduction |
| Optical, dielectric, and electric | High optical reflectivity <br> High skin charge density <br> High electricity (Harvesting) | |
| Chemical | High chemical reactivity <br> High catalytic activity | |
| Electron and phonon dynamics | Long lifetimes <br> Skin shell extra lifetimes | |

## 2. The coupling hydrogen bond and its scaling laws

The tetrahedrally-coordinated structure and the unique coupling HB foster water a homogeneous and ordered fluctuating crystal [62]. The identical members of electron lone pairs ":" and protons (H$^+$) reserve the configuration, number, and orientation of the HBs over wide ranges of temperature and pressure. Being the unique and efficient functional unit, HB relaxation in segmental length and energy determines uniquely performance of water under a load of perturbation.

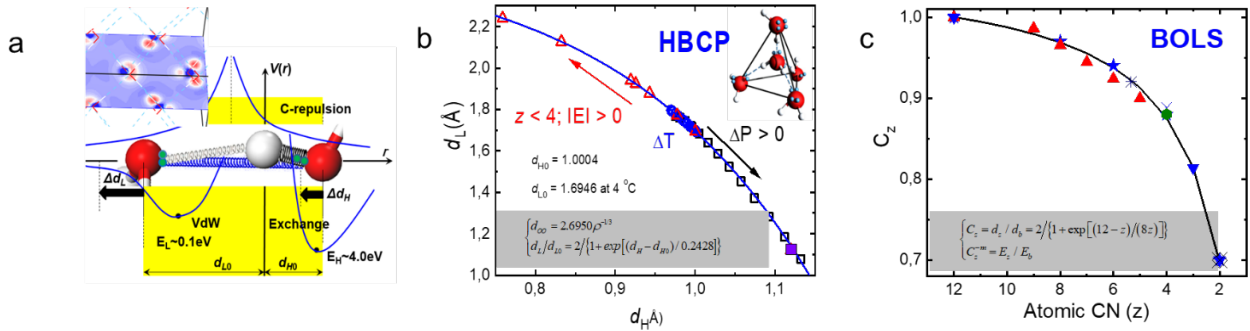

**Fig. 1 HB potentials, water structure, and scaling relations[62]**. (a) The potential contains the O:H attraction (0.1~0.2 eV, represented with van der Walls force), H-O polar covalent exchange interaction (4.0~5.0 eV), and O—O repulsive Coulomb coupling. (b) HB lengths cooperative relaxation. Inset shows that the H$_2$O:4H$_2$O motif of C$_{3v}$ symmetry. (c) atomic undercoordination-resolved bond contraction.

**Fig. 1**a illustrates the potential functions of the coupling system. The inter-oxygen repulsion endows the HB with an asymmetrical, coupled, and short-range oscillator pair HB that integrates the O:H and the H−O interactions. The potential contains the O:H attraction, H-O exchange interaction, and O—O Coulomb repulsion. Under any perturbation, oxygen anions travel along the HB in the same direction by different amounts with respect to the H coordination origin. The O—O separation changes proceed by lengthening one segment and shortening the other. The relaxation neither alters the nature of the HB nor does its orientation.

The inset in **Fig. 1**b illustrates the H$_2$O:4H$_2$O motif with oriented HBs of water. Water emits energy through H−O expansion and absorbs energy at inverse. Water dissipates energy through O:H relaxation and dissociation. The HB segmental length, energy, and inner angle form the parameters to feature the dynamics of boding and electrons as well as the performance of water.

**Fig. 1**b illustrates the HBCP notion regulating the HB segmental length cooperativity, and formulation [31]. The $d_{L0}$ = 1.6946 and $d_{H0}$ = 1.0004 Å are references for the tetrahedrally-coordinated structure of unity density at 277 K. Given a density, one can obtain the $d_H$ and $d_L$ values and their correlation Structural transition from the hcp to other alternatives should hold the correlation. For nanodroplets of 1.7-4.0 nm across, the skin H−O bond is 0.9 Å [86].

This relation indicates that under a load of any perturbation, such as electric field ($\varepsilon$), pressure (P), molecular undercoordination (z < 4), or temperature (T), the H-O ($d_H$) and O:H ($d_L$) segment length will shift cooperatively, associated with the electron lone pairs polarization toward or surpass the Fermi energy unless thermal fluctuation becomes dominance.

**Fig. 1**c describes the BOLS premise regulating the atomic CN dependence of the length $d_z$ and energy $E_z$ of a bond between z-coordinated atoms [87]. Where m is the bond nature index that correlates the bond energy to the bond length in the form of $E_z \propto d_z^{-m}$. The BOLS-NEP notion [87] indicates that a bond between fewer-coordinated atoms becomes shorter and stiffer. Bond contraction densifies electrons populated in the core and bonding orbitals, while bond stiffening deepens the potential that entraps electrons. These densely entrapped electrons, and meanwhile, polarize the edge atoms, turning them into dipoles that create topological states.

The spontaneous contraction only happens to bonds within the outermost two atomic spacings for a plane surface and could be more for nanocrystals to form the electron double layer [86, 88]. The NEP endows the undercoordinated atoms with exotic properties such as the single-atom catalytic activity and reactivity [89], edge superconductivity conductivity of topological insulators [35], and the high-$T_C$ superconductivity of monolayer [90], and the one-dimensional [91] cuprite oxides. Bond contraction in the skin of 2.13 bond length [86] stems from the size dependency of known properties such as bandgap, dielectric constant, mechanical strength, thermal stability, and magnetism [63]. The H-O bond of fewer-coordinated water molecules follows the BOLS predicted trend, which elongates and polarizes the O:H nonbonding electrons.

### 3.  Quasisolidity transiting Liquid to Ice

Water forms phases of XI, $I_c$, $I_h$, Quasisolid (QS), Liquid, and vapors associated with density oscillation when heated under the ambient pressure, which has been long standing puzzle sing 1611 when Galilei argued that ice floating results from density change. The HBCP premise allows us to solve this issue by introducing the segmental specific heat of Debye approximation. The specific heat depends on two factors: one is the cohesive energy E that equals the integral of the specific heat curve, and the other Deby temperature following Einstein relation, $\Theta_D \propto \omega$. With known $(\omega, E)_x$, one could readily define the specific heat $\eta_x$. The segmental disparity defines two specific heat curves and hence the phase boundaries and mass density variation with temperature [60].

**Fig. 2**a shows the disparate specific heats. The intersection defines an exotic phase between Liquid and Ice $I_h$, called quasisolid, within which mass density drops at cooling. The phase boundaries correspond to extreme mass density and close temperatures of melting $T_m$ and ice nucleation $T_N$. The $T_m$ and $T_N$ are adjustable by tuning the segmental frequency and cohesive energy. For instance, electrification or molecular undercoordination raises the $(E, \omega)_H$ for the H-O bond and lowers the $(E, \omega)_L$ for the O:H, which disperses outwardly the QS boundaries, as denoted in **Fig. 2**a. Approximately, the $T_m$ depends on the $E_h$ and the $T_N$ and $T_V$ on the $E_L$.

The density oscillation is correlated to the specific heat ratio in each phase, see **Fig. 2**b. The correspondence indicates that the segment of lower specific heat follows the regular rule of thermal expansion while other does contrastingly as a result of O—O coupling interaction. IN the QS phase, $\eta_H$ is lower than $\eta_L$, so the H-O bond experiences cooling contraction but the O:H expands. The H-O contracts less than the extent of O:H expansion, so volume expands, resulting in ice buoyancy. In the Liquid and ice, specific heat inversion alters the situation, resiting in thermal expansion of water and ice.

Fig. 2c and d show the conversion of the measured $\rho(T)$ into the $d_{O-O}(T)$ for water droplets of 1.4 and 4.4 nm sowing the outward disperse of the $T_m$ by taking the 277 K as the reference of calibration [4, 32]. The $d_{O-O}$ values of 2.70 Å measured at 298 K, and 2.71 Å at 256.5 K [92] match the derivative $\rho(T)$ and confirm the cooling expansion of the O—O distance in the quasisolid phase.

Therefore, the quasisolidity, the phase boundaries, and mass density variation result from the HBCP under thermal perturbation and the O—O repulsive coupling dictates all these consequences.

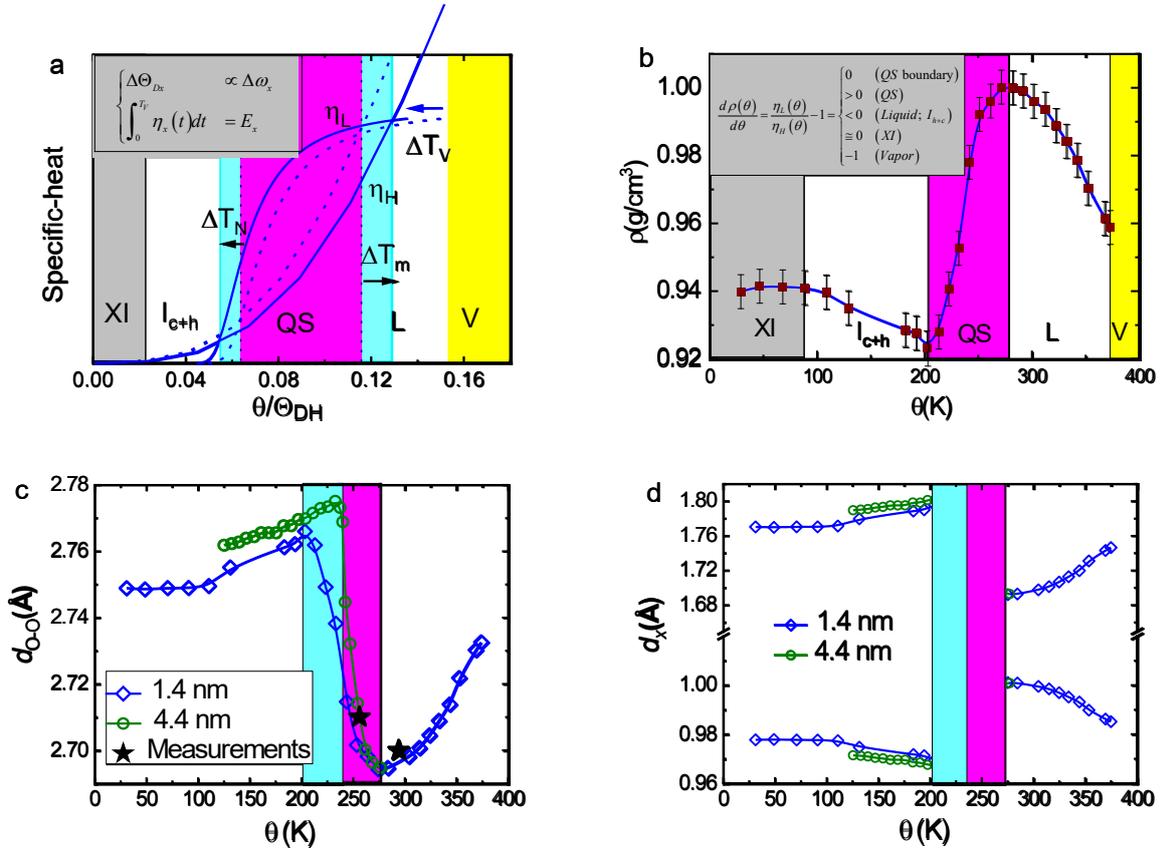

**Fig. 2 Specific heat disparity defined phase boundaries and oscillating mass density** [93]. (a) Vibration frequency $\omega_x$ and cohesive energy $E_x$ defines segmental specific heat $\eta_x$ and phase boundaries, and (b) mass density of water under ambient pressure. (c) The $d_{O-O}(T)$ and (d) the $d_x(T)$ converted from the measured density $\rho(T)$ for water droplets ($T < 273$ K) and bulk ($T > 273$ K) [4, 32]. The $d_{O-O}$ values 2.70 Å measured at 298 K, and 2.71 Å at 256.5 K [92] match the $d_{O-O}(T)$ profiles. Insets (a) and (b) formulates the specific heat and specific heat ratio resolved slope of density variation.

## 4. Supersolidity due to electrification and molecular undercoordination

We explore the performance of the skin HB using DFT calculations and solving the Lagrangian oscillation dynamics. **Fig. 3**a and b show the $(H_2O)_N$ cluster size and electric field dependence of the HB segmental lengths and potential paths. The O—O repulsion and molecular under-CN stretches the HB by shortening the H-O bond and lengthening the O:H nonbond. Electrification by applying an electric field has the same effect as molecular CN on the HB segmental lengths [61]. Lagrangian-Laplacian resolution to the HB oscillation dynamics converted the segmental lengths and vibration frequencies into their force constants and cohesive energies, giving rise to their relaxing potential paths (**Fig. 3**b).

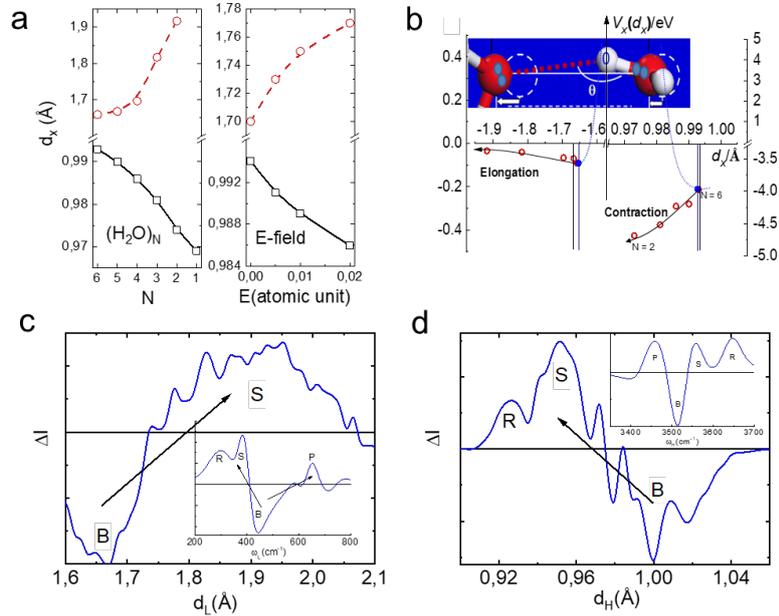

**Fig. 3 Droplet size and electric field resolved HB segmental lengths, potential paths, and vibration frequencies** [31]. (a) Molecular CN reduction and electrification effect the same on stretching O—O distance by shortening the H-O bond and lengthening the O:H nonbond, associated with polarization. (b) the relaxing potential paths of the HB. Transition of (c) the O:H nonbond and (d) the H–O bond from the bull (B) to the skin (S) and the H–O radicals (R) in HB segmental lengths and stiffness. The P components arise from calculation artifacts [35].

Further DFT computations on the skin effect of a unit cell containing 64 water molecules confirmed the expected relaxation of the segmental lengths and frequencies [35]. The PDPS in **Fig. 3c** and **d** refine the skin effect on the HB vibration frequencies and lengths. The refinement reveals the abundance, stiffness, and structure order transition of the H-O bonds from bulk (B) to skin (S) and free radicals (R). The $d_H$ contracts from 1.00 to 0.95 and 0.92 Å, as the H-O transits from the bulk to the skin and the dangling radicals. The $d_L$ elongates from 1.67 to a broad range centered at 1.90 Å of high fluctuation. The O:H frequencies $\omega_L$ turn out to be softer and the H-O frequencies $\omega_H$ become stiffer. The feature P arises from artifacts in DFT computations.

Wang and his team [94, 95] calculated the molecular site and (H$_2$O)$_{17-25}$ cluster size-resolved H-O stiffness. Results show consistently that the frequency reduces from 3760 cm$^{-1}$ for the dangling H-O bond to 2900 cm$^{-1}$ as moving inward to the center of the clusters. According to SFG spectroscopy [96], the H−O bonds pointing from the outermost layer of ice-I$_h$(001) surface inward vibrate at frequencies above 3270 cm$^{-1}$ while those from the second layer up are below. These observations confirmed the BOLS expectation on the H-O bonds [31] and their extreme sensitivity to the coordination conditions.

We now turn to prove the core-shell structured water droplet, as resolved from the size dependent phonon frequency shift of Si, CeO$_2$, and SnO$_2$ nanocrystals [86]. One can obtain the PDPS peaks by subtracting the specific H-O phonon peak collected from small emission angles from those of large emission angles upon all peaks area normalized [87]. **Fig. 4**a shows the PDPS refinement of H−O phonon transition in abundance, stiffness, and structure order from ice mode of 3150 cm$^{-1}$ and 3200 cm$^{-1}$ for water [97] to their skins. The skins of water ice share the 3450 cm$^{-1}$ vibration frequency, which evidences that a supersolid skin covers both water and ice instead of the historical perception of ice covering water or water wrapping ice. Ice skin is estimated 9/4 folds thick that of liquid water, as the latter is subjected to thermal fluctuation.

**Fig. 4**b shows the PDPS refinement of the phonon transition in abundance and frequency for the sized droplets of (5%D + 95%H) water with a focus on D-O vibration. The PDPS peak area resolves the fraction of phonons transiting from the mode of ordinary water to skins of water droplets (**Fig. 4**c): $F(R) = 0.27R^{-1}$, or the skin volume ratio to the entire nanosphere of R radius [87].

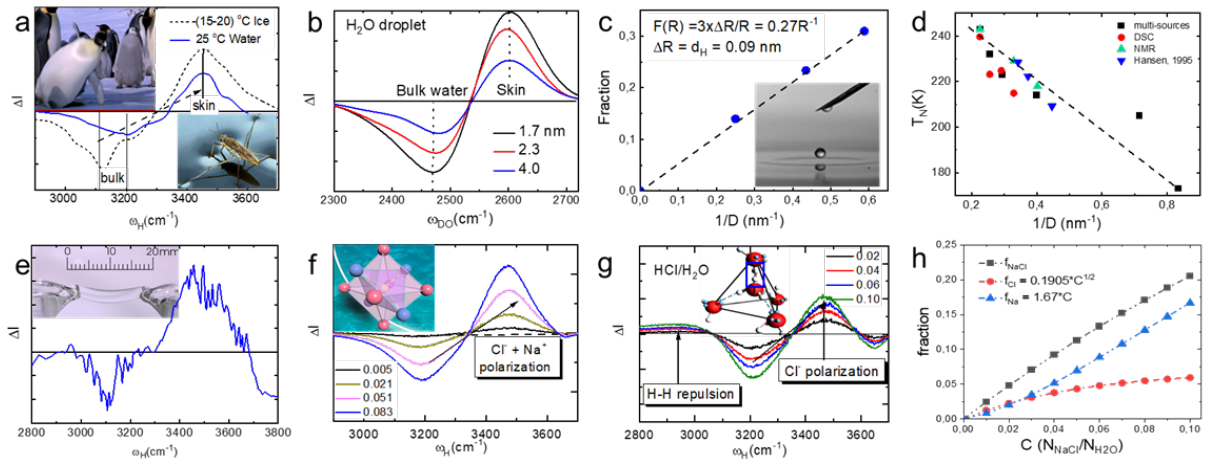

**Fig. 4  PDPS resolution of the supersolid phase.** (a) Skin-bulk resolved ω$_H$ transition from 3150/3200 cm$^{-1}$ for bulk ice/water [97] to their skin mode of 3450 cm$^{-1}$. (b) Core-shell resolved $\omega_D$ PDPS and (c) the fraction F(R) of skin phonon abundance [86]. (d) Size dependent droplets supercooling [32, 98]. Supersolidity of (e) water bridge (inset) [39] and ionic hydration cells of (f) NaCl and (g) HCl solutions. (h) Concentration dependence of the fraction F(C) of hydration cells phonon abundance[40].

One can readily evaluate the shell thickness ΔR of a droplet using the PDPS strategy, as shown in **Fig. 4**c. Given the volume of a sphere, V ∝ $R^3$, one can derive the skin volume ratio to the entire sphere, F(R) ∝ ΔV$_{skin}$/V = ΔN/N = 3ΔR$_{skin}$/R. The ΔR$_{skin}$ = F(R)R/3 = 0.90 Å is the thickness of the skin, which is identical to the H–O dangling radicals featured at 3610 cm$^{-1}$ [31]. The PDPS distills only the H-O bond in the outermost layer of the skin without discriminating the transition from the skin to the core. The skin D-O bond of 10% short corresponds to a 3.0 Å thick of the outermost layer. The D/H-O contraction and associated molecular polarization dictate the function of a nanodroplet.

**Fig. 4**d shows the droplet size dependence of supercooling for ice formation resolved using differential scanning calorimetry (DSC) and nuclear magnetron resonance spectroscopy (NMR) [32, 98]. The trend of T$_N$ depression depends on the skin volume ratio, ΔT$_N$(R) ∝ = F(R), which clarifies that the skin O:H nonbond softening inside the skin is responsible for nanodroplets supercooling for ice formation.

**Fig. 4**e shows the same supersolidity of water bridge induced by electrification. The inset displays the water bridge formed between two containers under a $10^6$ eV/μm field. The bridge has a rubber-like elasticity of around 6 MPa, which is stable till 330 K [37, 38]. Insets of **Fig. 4**f and g illustrate the three-dimensional (±; H$_3$O$^+$)·4H$_2$O:6H$_2$O hydration cells for ions and proton solvation[40] and the PDPS refinement of phonon spectral peaks. Ions occupy eccentrically the tetrahedrally-coordinated hollow sites to polarize their nearest and the next nearest neighbors to form the supersolid hydration cells.

Solvation dissolves HCl into Cl$^-$ and H$^+$. The anion Cl$^-$ performed the same as ions do in a salt solution that polarize and stretch the HB. Neither covalent nor ionic bonds could form between an ionic solute and its hydrating H$_2$O molecules. The proton joins then an H$_2$O to form the H$_3$O$^+$. The H$_3$O$^+$ replaces a water molecule and alters one HB into the repulsive H↔H in the form of O-H↔H-O. The H↔H repulsion lengthens the H-O of the solvent, according to the HBCP regulation.

PDPS refinement resolves the same H-O spectral features centered at 3450 cm$^{-1}$ for the water bridge and the ionic hydration cells in (Na, H)Cl solutions [40]. The H↔H repulsion lengthens the hydrating H-O bond, which turns out the board spectral hump of lower frequencies in **Fig. 4**g. The fraction coefficient f(C) of phonon abundance transition corresponds to the total volume of hydration cells in the concentrated solutions. The f(C)/C is proportional to the hydration cell size. **Fig. 4**h compares the f(C) for (Na, H)Cl solutions. The H↔H is incapable of polarization, and hence, $f_{HCl}(C) = f_{Cl}(V)$, and $f_{NaCl}(C) = f_{Na}(C) + f_{Cl}(C)$.

The linear form of $f_{Na}(C)$ indicates that the hydration cell volume $f_{Na}(C)/C$ of a Na$^+$ is independent of solute concentration as the hydrating H$_2$O molecules fully screen the electric field of the small Na$^+$. The nonlinear C$^{1/2}$ form of Cl$^-$ indicates the involvement of inter-anion repulsion that weakens the local electric field. The $f_{Cl}(C)/C \propto C^{-1/2}$ as the hydrating molecules can only partly screen the large Cl$^-$. The supersolidity of the NaCl solution is responsible for the solution viscosity, formulated by Jones-Dole in 1929 [99]: $\eta(C) = aC + bC^{1/2}$, with a and b being adjustable coefficients. One can reformulate the viscosity as $\eta(C) = af_{Na}(C) + bf_{Cl}(C)$, other than contributing from inter-solute and solute-solvent interactions as Jones-Dole deemed. The resultant of the screened polarization of ions and the inter-anion repulsion dictates the viscosity of alkali-halide solutions.

5. **Electronic signatures, electron and phonon lifetimes**

According to the BOLS-NEP premise, when atoms are under-coordinated, the bond strength gain shifts the core level even deeper and shifts up the nonbonding electrons in energy. Energy band theory suggests that the O 1s level shift be proportional to the H-O bond energy, as the O:H contribution is negligible. XPS profile in **Fig. 5**a shows that the O 1s level shifts from 536.6 to 538.1 and even 539.7 eV when an H$_2$O transits its location from bulk to the skin and the isolated gaseous monomer [100, 101].

A free electron can serve as the probe for the local environment when injected into water without hanging the solvent's geometry. The locally oriented H$_2$O molecules entrap the hydrated electron by forming a (e$^-$)·4H$_2$O motif, which inverses the polarity of the (Na$^+$)·4H$_2$O hydration cell. The intensity of the entrapment varies from site to site because of the polarization extent of the coordination resolved hydrating H2O dipoles. The bound energy of the hydrated electron features the energy required to eject an electron out of the liquid.

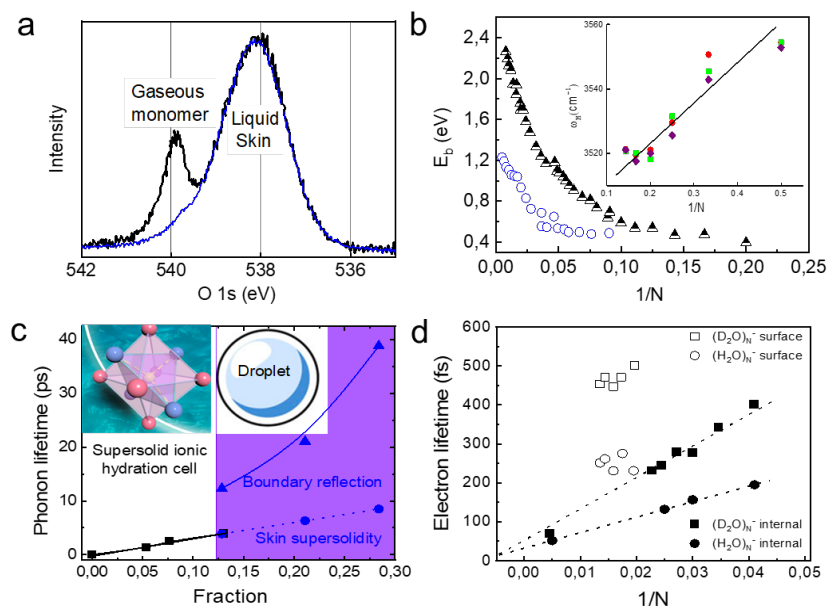

**Fig. 5 Site resolved energies and lifetimes of electrons phonons.** (a) O 1s level shifts from 536.6 (unseen) to 538.1 and 540 eV when the H2O molecule transmits from bulk to the skin and gaseous monomer [101]. (b) $(H_2O)_N$ cluster size dependent shift of the $\omega_H$ (inset) and the bound energy of electrons hydrated in the cluster core (for the initial 2.4 eV) and skin (1.2 eV) till 0.4 eV for N = 5 [94, 102]. (c) Comparison of the D−O phonon lifetimes of the sized droplet and the concentrated NaBr solutions [74], and (d) molecular site and cluster size-resolved electron lifetimes [102]. The inset (c) illustrates the ionic hydration cell and the core-shelled droplet of D/H water.

Ultrafast pump-probe liquid-jet UPS revealed that the bound energy of an electron hydrated inside the core interior of droplets is centered at 2.4 eV and 1.2 eV when located inside the skin. The bound energies drop with the $(H_2O)_N$ size till 0.4 eV for N = 5. The $(H_2O)_N$ size dependence of the H−O phonon frequency shown in the inset of **Fig. 5**b demonstrates the undercoordination stiffens the H-O phonons whose frequency shift depends on the reciprocal of cluster size – the volume ratio of undercoordinated molecules.

Skin H−O bonds are even shorter and stiffer because the high surface curvature of smaller droplets lowers the molecular CN, which deepens the O 1s energy level and enhances polarization. Results of quantum computations [35, 94, 95], phonon and electron spectrometrics confirm consistently the HB between undercoordinated skin molecules follows the HBCP and BOLS-NEP regulations - H-O bond contraction and O:H nonbond elongation, O 1s core level self-entrapment and nonbonding electron polarization.

Ultrafast spectroscopy probes the coordination environment of a specific site by measuring the electron and phonon lifetimes or abundance dissipation dynamics. The phonon lifetime varies with the vibration frequency [103], and the electron lifetime is proportional to its inverse bound energy [102]. Both lifetimes depend positively on the extent of local HB polarization, decaying at a slower rate in the supersolid phase.

**Fig. 5**c displays the lifetimes of D-O phonons inside the sized water droplets and the NaBr solutions of different concentrations [74]. The phonon lifetime increases from 2.6 ps for pristine water to 3.9 and 6.7 ps for solutions of 32 and 8 $H_2O$ per NaBr solute, respectively. The phonon lifetime depends linearly on the total volume of supersolid hydration cells of the solution. In contrast, the phonon lifetime increases from 2.6 ps to 18 and 50 ps for droplets of 4.0 to 1.7 nm in size [74]. The much longer phonon lifetime contains components of the skin supersolidity and the boundary reflection [86] that forms quasi-standing waves.

**Fig. 5**d displays the droplet size and molecular site resolved electron lifetimes. The lifetimes of hydrated electrons increase with the drop of cluster size or the extent of supersolidity. The electron lives 100 ps longer and remains relatively stable in the skin than those within the droplets. The supersolidity of the NaBr hydration cell and the droplet skin slows down the dynamics of both phonons and the hydrated nonbonding electrons, in addition to the boundary reflection for the oscillating waves.

## 6. Discrimination between quasisolidity and supersolidity

One can verify the quasisolidity and supersolidity by aging samples in the constant temperature ambient[104]. Putting deionized water and saturated NaCl solution of the same volume into the thermally programmed chamber set the temperature initially at 294 K for 1h and lowers the temperature to and retained 254 K for 7h. The solution saturates at $N_{NaCl}/N_{H2O}$ =1/10 ratio, or 10 ten $H_2O$ molecules hydrate one pair of $Na^+$ and $Cl^-$ ions and that the solution contains no free $H_2O$. Thermal couples have amounted to the container and the detected samples to monitor the instantaneous temperatures.

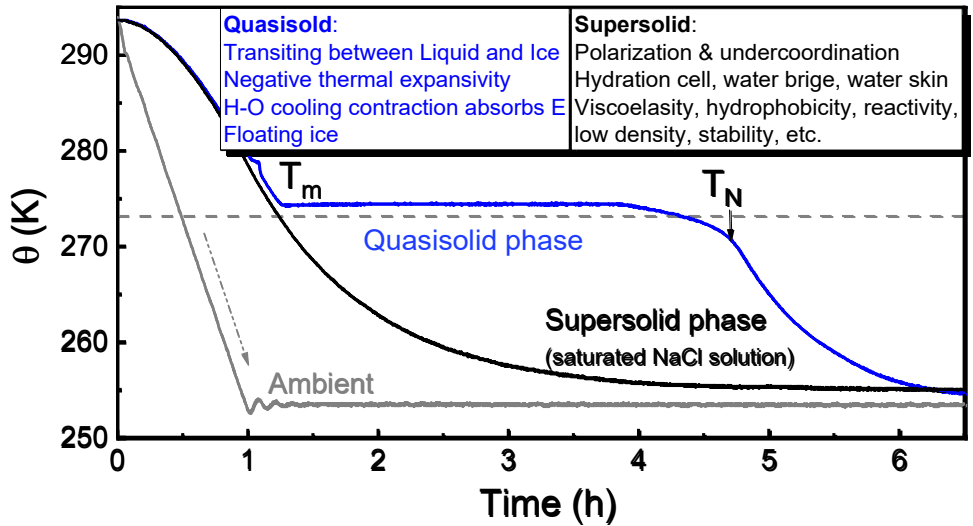

Fig. 6 The θ(t) decay lines revealed Liquid-QS-Ice transition and QS phase energy absorption for water but not for NaCl supersolid solution [104].

Fig. 6 discriminates the thermal θ(t) decay lines of water and the salt solution. Water undergoes the Liquid-QS-Ice transition, presenting the $T_m$ and $T_N$, corresponding to the second derivatives of the $θ''(t) = 0$. In the $t(T_N)$- $t(T_m)$ period, or the QS phase, the H-O bond absorbs energy by cooling contraction. The plateau area is proportional to the heat absorbed. Outside the $t(T_N)$- $t(T_m)$ period, the H-O bond expands cooling expansion instead to emit energy that heats the sample, deviating the θ(t) from the temperature of the container. The sample temperature drops exponentially with decay time. This finding clarifies the bonding mechanism of the enthalpy of phase transition that may emits or absorbs energy. Bond contraction or formation absorbs energy but expansion or dissociation does it inverse.

However, the θ(t) decay for the supersolid solution extends the exponential forms of θ(t) for water without any sign of phase transition or energy absorption. This observation indicates that the saturated NaCl solution retains supersolidity without freezing at a temperature above 254 K. Therefore, the θ(t) decay profiles further evidence the presence of the QS phase of water and supersolidity for a saturated ionic solution.

The supersolidity has the consequence of polarization that disperses outwardly the QS phase boundary

- raising the $T_m$ and lowering $T_N$ and turning the quasisolidity into the supersolidity in performance.

## 7. Conclusion

Consistent evidence shows that water and droplets prefer the core-shell structural configuration. The HB follows the HBCP and the BOLS-NEP regulations. The disparity of the O:H-O bond in segmental length, energy, and specific heat foster the QS phase of negative thermal expansivity. The outmost skin shell of water droplets is 0.3 nm thick. Electrification or molecular undercoordination stretches and polarizes the O—O by ~8% lengthening the O:H soft oscillators by up to ~18% and shortening the H-O bond by ~10%, which lowers 2~5% the mass density. The polarization results in high hydrophobicity, 4.8% more surface charge, mechanical strength, thermal stability and diffusivity, and optical reflectivity. The skin elastic O:H soft phonons and polarized $H_2O$ dipoles furnished the skins of water and droplets with catalytic activity, chemical reactivity, electricity, elastic adaptivity, and superlubricity, and salt solutions the Hofmeister effect – protein solubility. Thermal decay measurements of deionized water and saturated NaCl solution have confirmed the QS and supersolidity of water, clarifying the bonding origin of the enthalpy of phase transition. Further extension of the knowledge to other systems involving electron lone pairs or the coupling HB under electrification and molecular undercoordination would be even more fascinating and promising.

**Declaration of competing interest**

No competing financial interests or personal relationships is declared.

**Acknowledgments**

Financial support from the National Natural Science Foundation of China (Nos: 12150001 (BW)) is gratefully acknowledged.

**References**

[1] G. Malenkov, Liquid water and ices: understanding the structure and physical properties, J. Phys.-Condes. Matter, 21 (2009) 283101.
[2] J.L. Green, D.J. Durben, G.H. Wolf, C.A. Angell, Water and Solutions at Negative Pressure: Raman Spectroscopic Study to -80 Megapascals, Science, 249 (1990) 649-652.
[3] T. Head-Gordon, M.E. Johnson, Tetrahedral structure or chains for liquid water, Proc Natl Acad Sci U S A, 103 (2006)

7973-7977.
[4] F. Mallamace, C. Branca, M. Broccio, C. Corsaro, C.Y. Mou, S.H. Chen, The anomalous behavior of the density of water in the range 30 K < T < 373 K, Proc Natl Acad Sci U S A, 104 (2007) 18387-18391.
[5] F. Mallamace, M. Broccio, C. Corsaro, A. Faraone, D. Majolino, V. Venuti, L. Liu, C.Y. Mou, S.H. Chen, Evidence of the existence of the low-density liquid phase in supercooled, confined water, Proc Natl Acad Sci U S A, 104 (2007) 424-428.
[6] E. Pennisi, Water's Tough Skin, Science, 343 (2014) 1194-1197.
[7] G. Amit, Why is ice slippery?, New Scientist, London, 2015, pp. 38.
[8] R. Rosenberg, Why ice is slippery?, Physics Today, DOI (2005) 50-55.
[9] L. Canale, J. Comtet, A. Niguès, C. Cohen, C. Clanet, A. Siria, L. Bocquet, Nanorheology of interfacial water during ice gliding, Physical Review X, 9 (2019) 041025.
[10] X.F. Gao, L. Jiang, Water-repellent legs of water striders, Nature, 432 (2004) 36-36.
[11] C.Q. Sun, Thermo-mechanical behavior of low-dimensional systems: The local bond average approach, Progress in Materials Science, 54 (2009) 179-307.
[12] P. Xu, B. Cui, Y. Bu, H. Wang, X. Guo, P. Wang, Y.R. Shen, L. Tong, Elastic ice microfibers, Science, 373 (2021) 187-192.
[13] E.M. Schulson, A flexible and springy form of ice, Science, 373 (2021) 158-158.
[14] S.H. Oh, J.G. Han, J.-M. Kim, Long-term stability of hydrogen nanobubble fuel, Fuel, 158 (2015) 399-404.
[15] X. Zhang, X. Liu, Y. Zhong, Z. Zhou, Y. Huang, C.Q. Sun, Nanobubble Skin Supersolidity, Langmuir, 32 (2016) 11321-11327.
[16] D. Shin, J. Hwang, W. Jhe, Ice-VII-like molecular structure of ambient water nanomeniscus, Nature communications, 10 (2019) 1-8.
[17] P. Miranda, L. Xu, Y. Shen, M. Salmeron, Icelike water monolayer adsorbed on mica at room temperature, Physical review letters, 81 (1998) 5876.
[18] K. Meister, S. Strazdaite, A.L. DeVries, S. Lotze, L.L. Olijve, I.K. Voets, H.J. Bakker, Observation of ice-like water layers at an aqueous protein surface, Proceedings of the National Academy of Sciences, 111 (2014) 17732-17736.
[19] Z. He, J. Zhou, X. Lu, B. Corry, Ice-like water structure in carbon nanotube (8, 8) induces cationic hydration enhancement, The Journal of Physical Chemistry C, 117 (2013) 11412-11420.
[20] C. Wang, H. Lu, Z. Wang, P. Xiu, B. Zhou, G. Zuo, R. Wan, J. Hu, H. Fang, Stable Liquid Water Droplet on a Water Monolayer Formed at Room Temperature on Ionic Model Substrates, Physical Review Letters, 103 (2009) 137801-137804.
[21] R. Helmy, Y. Kazakevich, C.Y. Ni, A.Y. Fadeev, Wetting in hydrophobic nanochannels: A challenge of classical capillarity, Journal of the American Chemical Society, 127 (2005) 12446-12447.
[22] M. Mezger, H. Reichert, S. Schoder, J. Okasinski, H. Schroder, H. Dosch, D. Palms, J. Ralston, V. Honkimaki, High-resolution in situ x-ray study of the hydrophobic gap at the water-octadecyl-trichlorosilane interface, Proc Natl Acad Sci U S A, 103 (2006) 18401-18404.
[23] A. Uysal, M. Chu, B. Stripe, A. Timalsina, S. Chattopadhyay, C.M. Schlepütz, T.J. Marks, P. Dutta, What x rays can tell us about the interfacial profile of water near hydrophobic surfaces, Physical Review B, 88 (2013) 035431.
[24] X. Zhang, Y. Huang, Z. Ma, L. Niu, C.Q. Sun, From ice supperlubricity to quantum friction: Electronic repulsivity and phononic elasticity, Friction, 3 (2015) 294-319.
[25] J.H. Weijs, D. Lohse, Why Surface Nanobubbles Live for Hours, Physical Review Letters, 110 (2013) 054501.
[26] L.-J. Zhang, J. Wang, Y. Luo, H.-P. Fang, J. Hu, A novel water layer structure inside nanobubbles at room temperature Nuclear Science and Techniques, 25 (2014) 060503
[27] C. Chen, J. Li, X. Zhang, The existence and stability of bulk nanobubbles: a long-standing dispute on the experimentally observed mesoscopic inhomogeneities in aqueous solutions, Communications in Theoretical Physics, 72 (2020) 037601.
[28] Y. Zhou, Y. Zhong, Y. Gong, X. Zhang, Z. Ma, Y. Huang, C.Q. Sun, Unprecedented thermal stability of water supersolid skin, Journal of Molecular Liquids, 220 (2016) 865-869.
[29] E.B. Mpemba, D.G. Osborne, Cool?, Physics Education, 14 (1979) 410-413.
[30] B. Slater, A. Michaelides, Surface premelting of water ice, Nature Reviews Chemistry, 3 (2019) 172-188.
[31] Y. Huang, X. Zhang, Z. Ma, Y. Zhou, W. Zheng, J. Zhou, C.Q. Sun, Hydrogen-bond relaxation dynamics: Resolving mysteries of water ice, Coordination Chemistry Reviews, 285 (2015) 109-165.
[32] M. Erko, D. Wallacher, A. Hoell, T. Hauss, I. Zizak, O. Paris, Density minimum of confined water at low temperatures: a combined study by small-angle scattering of X-rays and neutrons, Phys Chem Chem Phys, 14 (2012) 3852-3858.
[33] M. James, T.A. Darwish, S. Ciampi, S.O. Sylvester, Z.M. Zhang, A. Ng, J.J. Gooding, T.L. Hanley, Nanoscale condensation of water on self-assembled monolayers, Soft Matter, 7 (2011) 5309-5318.
[34] H. Qiu, W. Guo, Electromelting of Confined Monolayer Ice, Physical Review Letters, 110 (2013) 195701.
[35] X. Zhang, Y. Huang, Z. Ma, Y. Zhou, W. Zheng, J. Zhou, C.Q. Sun, A common supersolid skin covering both water and ice, Phys Chem Chem Phys, 16 (2014) 22987-22994.

[36] D.P. Singh, J.P. Singh, Delayed freezing of water droplet on silver nanocolumnar thin film, Applied Physics Letters, 102 (2013) 243112.
[37] A.A. Aerov, Why the water bridge does not collapse, Physical Review E, 84 (2011) 036314.
[38] J. Woisetschlager, K. Gatterer, E.C. Fuchs, Experiments in a floating water bridge, Experiments in Fluids, 48 (2010) 121-131.
[39] R. Ponterio, M. Pochylski, F. Aliotta, C. Vasi, M. Fontanella, F. Saija, Raman scattering measurements on a floating water bridge, Journal of Physics D: Applied Physics, 43 (2010) 175405.
[40] C.Q. Sun, Aqueous charge injection: solvation bonding dynamics, molecular nonbond interactions, and extraordinary solute capabilities, International Reviews in Physical Chemistry, 37 (2018) 363-558.
[41] J.K. Lee, K.L. Walker, H.S. Han, J. Kang, F.B. Prinz, R.M. Waymouth, H.G. Nam, R.N. Zare, Spontaneous generation of hydrogen peroxide from aqueous microdroplets, Proc Natl Acad Sci U S A, 116 (2019) 19294-19298.
[42] M.A. Mehrgardi, M. Mofidfar, R.N. Zare, Sprayed water microdroplets are able to generate hydrogen peroxide spontaneously, Journal of the American Chemical Society, 144 (2022) 7606-7609.
[43] D. Xing, Y. Meng, X. Yuan, S. Jin, X. Song, R.N. Zare, X. Zhang, Capture of hydroxyl radicals by hydronium cations in water microdroplets, Angewandte Chemie, 134 (2022) e202207587.
[44] Y. Meng, R.N. Zare, E. Gnanamani, One-Step, Catalyst-Free Formation of Phenol from Benzoic Acid Using Water Microdroplets, Journal of the American Chemical Society, 145 (2023) 19202-19206.
[45] X. Song, C. Basheer, R.N. Zare, Water Microdroplets-Initiated Methane Oxidation, Journal of the American Chemical Society, 145 (2023) 27198–27204.
[46] X. Song, C. Basheer, Y. Xia, J. Li, I. Abdulazeez, A.A. Al-Saadi, M. Mofidfar, M.A. Suliman, R.N. Zare, One-step Formation of Urea from Carbon Dioxide and Nitrogen Using Water Microdroplets, Journal of the American Chemical Society, 145 (2023) 25910–25916.
[47] X. Song, C. Basheer, R.N. Zare, Making ammonia from nitrogen and water microdroplets, Proceedings of the National Academy of Sciences, 120 (2023) e2301206120.
[48] S. Narayan, J. Muldoon, M. G. Finn, V. V. Fokin, H.C. Kolb, K.B. Sharpless, "On water": Unique reactivity of organic compounds in aqueous suspension, Angewandte Chemie International Edition, 44 (2005) 3275–3279.
[49] M.F. Ruiz-Lopez, J.S. Francisco, M.T. Martins-Costa, J.M. Anglada, Molecular reactions at aqueous interfaces, Nature Reviews Chemistry, 4 (2020) 459-475.
[50] H. Hao, I. Leven, T. Head-Gordon, Can electric fields drive chemistry for an aqueous microdroplet?, Nature communications, 13 (2022) 280.
[51] J. Nauruzbayeva, Z. Sun, A. Gallo Jr, M. Ibrahim, J.C. Santamarina, H. Mishra, Electrification at water–hydrophobe interfaces, Nature communications, 11 (2020) 5285.
[52] S. Lin, X. Chen, Z.L. Wang, Contact electrification at the liquid–solid interface, Chemical Reviews, 122 (2021) 5209-5232.
[53] X. Wang, F. Lin, X. Wang, S. Fang, J. Tan, W. Chu, R. Rong, J. Yin, Z. Zhang, Y. Liu, Hydrovoltaic technology: from mechanism to applications, Chemical Society Reviews, 51 (2022) 4902-4927.
[54] W. Xu, H. Zheng, Y. Liu, X. Zhou, C. Zhang, Y. Song, X. Deng, M. Leung, Z. Yang, R.X. Xu, A droplet-based electricity generator with high instantaneous power density, Nature, 578 (2020) 392-396.
[55] L. Wang, Y. Song, W. Xu, W. Li, Y. Jin, S. Gao, S. Yang, C. Wu, S. Wang, Z. Wang, Harvesting energy from high‐frequency impinging water droplets by a droplet‐based electricity generator, EcoMat, 3 (2021) e12116.
[56] K.R. Wilson, R.D. Schaller, D.T. Co, R.J. Saykally, B.S. Rude, T. Catalano, J.D. Bozek, Surface relaxation in liquid water and methanol studied by x-ray absorption spectroscopy, J. Chem. Phys., 117 (2002) 7738-7744.
[57] M. Faraday, Note on Regelation, Proceedings of the Royal Society of London, 10 (1859) 440-450.
[58] J. Thomson, Note on Professor Faraday's Recent Experiments on Regelation, Proceedings of the Royal Society of London, 10 (1859) 151-160.
[59] T. Hynninen, V. Heinonen, C.L. Dias, M. Karttunen, A.S. Foster, T. Ala-Nissila, Cutting ice: nanowire regelation, Physical review letters, 105 (2010) 086102.
[60] C.Q. Sun, X. Zhang, X. Fu, W. Zheng, J.-l. Kuo, Y. Zhou, Z. Shen, J. Zhou, Density and phonon-stiffness anomalies of water and ice in the full temperature range, Journal of Physical Chemistry Letters, 4 (2013) 3238-3244.
[61] C.Q. Sun, Water electrification: Principles and applications, Advances in Colloid and Interface Science, 282 (2020) 102188.
[62] C.Q. Sun, Y. Huang, X. Zhang, Z. Ma, B. Wang, The physics behind water irregularity, Physics Reports, 998 (2023) 1-68.
[63] C.Q. Sun, Size dependence of nanostructures: Impact of bond order deficiency, Progress in Solid State Chemistry, 35 (2007) 1-159.
[64] D.Y. Kim, M.H.W. Chan, Upper limit of supersolidity in solid helium, Physical Review B, 90 (2014) 064503.
[65] G. Gonella, E.H. Backus, Y. Nagata, D.J. Bonthuis, P. Loche, A. Schlaich, R.R. Netz, A. Kühnle, I.T. McCrum, M.T. Koper, Water at charged interfaces, Nature Reviews Chemistry, 5 (2021) 466-485.

[66] E.C. Fuchs, A.D. Wexler, A.H. Paulitsch-Fuchs, L.L. Agostinho, D. Yntema, J. Woisetschläger, The Armstrong experiment revisited, The European Physical Journal Special Topics, 223 (2014) 959-977.
[67] Y. Bronstein, P. Depondt, L.E. Bove, R. Gaal, A.M. Saitta, F. Finocchi, Quantum versus classical protons in pure and salty ice under pressure, Physical Review B, 93 (2016) 024104.
[68] M. Nagasaka, H. Yuzawa, N. Kosugi, Interaction between water and alkali metal ions and its temperature dependence revealed by oxygen K-edge x-ray absorption spectroscopy, The Journal of Physical Chemistry B, 121 (2017) 10957-10964.
[69] J. Chen, K. Nagashima, K.-i. Murata, G. Sazaki, Quasi-liquid layers can exist on polycrystalline ice thin films at a temperature significantly lower than on ice single crystals, Crystal Growth & Design, 19 (2018) 116-124.
[70] S. Toda, Y. Asakawa, Studies on the improvement of fuel combustion and vapour evaporation of small steam boiler: Effect of high voltage, Bulletin of the College of Agriculture and Veterinary Medicine Nihon University, DOI (1976).
[71] K. Ando, M. Arakawa, A. Terasaki, Freezing of micrometer-sized liquid droplets of pure water evaporatively cooled in a vacuum, Phys Chem Chem Phys, 20 (2018) 28435-28444.
[72] A. Agarwal, W.J. Ng, Y. Liu, Principle and applications of microbubble and nanobubble technology for water treatment, Chemosphere, 84 (2011) 1175-1180.
[73] S. Park, M.D. Fayer, Hydrogen bond dynamics in aqueous NaBr solutions, Proceedings of the National Academy of Sciences, 104 (2007) 16731-16738.
[74] S. Park, D.E. Moilanen, M.D. Fayer, Water Dynamics: The Effects of Ions and Nanoconfinement, The Journal of Physical Chemistry B, 112 (2008) 5279-5290.
[75] J. Verlet, A. Bragg, A. Kammrath, O. Cheshnovsky, D. Neumark, Observation of large water-cluster anions with surface-bound excess electrons, Science, 307 (2005) 93-96.
[76] Trainoff S, P. N., water droplet dancing on water surfaces, 2009.
[77] I.S. Klyuzhin, F. Ienna, B. Roeder, A. Wexler, G.H. Pollack, Persisting water droplets on water surfaces, The Journal of Physical Chemistry B, 114 (2010) 14020-14027.
[78] D. Wang, Y. Tian, L. Jiang, Abnormal Properties of Low‐Dimensional Confined Water, Small, 17 (2021) 2100788.
[79] C.-S. Zha, R.J. Hemley, S.A. Gramsch, H.-k. Mao, W.A. Bassett, Optical study of $H_2O$ ice to 120 GPa: Dielectric function, molecular polarizability, and equation of state, The Journal of chemical physics, 126 (2007).
[80] X. Zhang, T. Yan, Y. Huang, Z. Ma, X. Liu, B. Zou, C.Q. Sun, Mediating relaxation and polarization of hydrogen-bonds in water by NaCl salting and heating, Phys Chem Chem Phys, 16 (2014) 24666-24671.
[81] L.E. Bove, R. Gaal, Z. Raza, A.-A. Ludl, S. Klotz, A.M. Saitta, A.F. Goncharov, P. Gillet, Effect of salt on the H-bond symmetrization in ice, Proceedings of the National Academy of Sciences, 112 (2015) 8216-8220.
[82] D.T. Bregante, M.C. Chan, J.Z. Tan, E.Z. Ayla, C.P. Nicholas, D. Shukla, D.W. Flaherty, The shape of water in zeolites and its impact on epoxidation catalysis, Nature Catalysis, 4 (2021) 797-808.
[83] L. Zhou, X. Wang, H.-J. Shin, J. Wang, R. Tai, X. Zhang, H. Fang, W. Xiao, L. Wang, C. Wang, Ultra-high Density of Gas Molecules Confined in Surface Nanobubbles in Ambient Water, Journal of the American Chemical Society, DOI: 10.1021/jacs.9b11303 (2020).
[84] C. Gong, D. Li, X. Li, D. Zhang, D. Xing, L. Zhao, X. Yuan, X. Zhang, Spontaneous reduction-induced degradation of viologen compounds in water microdroplets and its inhibition by host–guest complexation, Journal of the American Chemical Society, 144 (2022) 3510-3516.
[85] A.J. Colussi, Mechanism of hydrogen peroxide formation on sprayed water microdroplets, Journal of the American Chemical Society, 145 (2023) 16315–16317.
[86] X. Yang, C. Peng, L. Li, M. Bo, Y. Sun, Y. Huang, C.Q. Sun, Multifield-resolved phonon spectrometrics: structured crystals and liquids, Progress in Solid State Chemistry, 55 (2019) 20-66.
[87] X.J. Liu, M.L. Bo, X. Zhang, L. Li, Y.G. Nie, H. Tian, Y. Sun, S. Xu, Y. Wang, W. Zheng, C.Q. Sun, Coordination-resolved electron spectrometrics, Chemical Reviews, 115 (2015) 6746-6810.
[88] W.J. Huang, R. Sun, J. Tao, L.D. Menard, R.G. Nuzzo, J.M. Zuo, Coordination-dependent surface atomic contraction in nanocrystals revealed by coherent diffraction, Nature Materials, 7 (2008) 308-313.
[89] J. Lin, A. Wang, B. Qiao, X. Liu, X. Yang, X. Wang, J. Liang, J. Li, J. Liu, T. Zhang, Remarkable Performance of Ir1/FeOx Single-Atom Catalyst in Water Gas Shift Reaction, Journal of the American Chemical Society, 135 (2013) 15314-15317.
[90] Y. Yu, L. Ma, P. Cai, R. Zhong, C. Ye, J. Shen, G.D. Gu, X.H. Chen, Y. Zhang, High-temperature superconductivity in monolayer Bi2Sr2CaCu2O8+δ, Nature, 575 (2019) 156-163.
[91] Z. Chen, Y. Wang, S.N. Rebec, T. Jia, M. Hashimoto, D. Lu, B. Moritz, R.G. Moore, T.P. Devereaux, Z.-X. Shen, Anomalously strong near-neighbor attraction in doped 1D cuprate chains, Science, 373 (2021) 1235-1239.
[92] U. Bergmann, A. Di Cicco, P. Wernet, E. Principi, P. Glatzel, A. Nilsson, Nearest-neighbor oxygen distances in liquid water and ice observed by x-ray Raman based extended x-ray absorption fine structure, The Journal of Chemical Physics, 127 (2007) 174504.
[93] Y. Huang, X. Zhang, Z. Ma, Y. Zhou, J. Zhou, W. Zheng, C.Q. Sun, Size, separation, structure order, and mass density of molecules packing in water and ice, http://www.nature.com/srep/2013/131021/srep03005/metrics, 3 (2013) 3005.

[94] B. Wang, W. Jiang, Y. Gao, Z. Zhang, C. Sun, F. Liu, Z. Wang, Energetics competition in centrally four-coordinated water clusters and Raman spectroscopic signature for hydrogen bonding, RSC Advances, 7 (2017) 11680-11683.
[95] Z. Zhang, D. Li, W. Jiang, Z. Wang, The electron density delocalization of hydrogen bond systems, Advances in Physics: X, 3 (2018) 1428915.
[96] Y. Otsuki, T. Sugimoto, T. Ishiyama, A. Morita, K. Watanabe, Y. Matsumoto, Unveiling subsurface hydrogen-bond structure of hexagonal water ice, Physical Review B, 96 (2017) 115405.
[97] T.F. Kahan, J.P. Reid, D.J. Donaldson, Spectroscopic probes of the quasi-liquid layer on ice, Journal of Physical Chemistry A, 111 (2007) 11006-11012.
[98] S. Jähnert, F.V. Chávez, G. Schaumann, A. Schreiber, M. Schönhoff, G. Findenegg, Melting and freezing of water in cylindrical silica nanopores, Phys Chem Chem Phys, 10 (2008) 6039-6051.
[99] G. Jones, M. Dole, The viscosity of aqueous solutions of strong electrolytes with special reference to barium chloride, Journal of the American Chemical Society, 51 (1929) 2950-2964.
[100] K. Nishizawa, N. Kurahashi, K. Sekiguchi, T. Mizuno, Y. Ogi, T. Horio, M. Oura, N. Kosugi, T. Suzuki, High-resolution soft X-ray photoelectron spectroscopy of liquid water, Phys Chem Chem Phys, 13 (2011) 413-417.
[101] B. Winter, E.F. Aziz, U. Hergenhahn, M. Faubel, I.V. Hertel, Hydrogen bonds in liquid water studied by photoelectron spectroscopy, J. Chem. Phys., 126 (2007) 124504.
[102] K.R. Siefermann, Y. Liu, E. Lugovoy, O. Link, M. Faubel, U. Buck, B. Winter, B. Abel, Binding energies, lifetimes and implications of bulk and interface solvated electrons in water, Nature Chemistry, 2 (2010) 274-279.
[103] S.T. van der Post, C.S. Hsieh, M. Okuno, Y. Nagata, H.J. Bakker, M. Bonn, J. Hunger, Strong frequency dependence of vibrational relaxation in bulk and surface water reveals sub-picosecond structural heterogeneity, Nature communications, 6 (2015) 8384.
[104] Y.J. Shen, X. Wei, Y. Wang, Y.T. Shen, L. Li, Y. Huang, K. Ostricov, C.Q. Sun, Energy absorbancy and freezing-temperature tunability of NaCl solutions during ice formation, J. Mol. Liq., 344 (2021) 117928.